\documentclass[a4paper,12pt,eqno]{article}

\usepackage{amsthm}
\usepackage{amssymb}
\usepackage[dvips]{graphicx}

\newtheorem{thm}{THEOREM}
\newtheorem{lem}[thm]{LEMMA}
\newtheorem{cor}[thm]{COROLLARY}

\newcommand{\ket}[1]{|#1\rangle}

\title{{\Large {\bf A new type of limit theorems for the one-dimensional quantum random walk}
}}
\author{{\small By Norio KONNO } \\
}

\date{\empty }
\begin{document}
\maketitle

\par\noindent
\begin{small}
{\bf Abstract}. In this paper we consider the one-dimensional quantum random walk $X^{\varphi} _n$ at time $n$ starting from initial qubit state $\varphi$ determined by $2 \times 2$ unitary matrix $U$. We give a combinatorial expression for the characteristic function of $X^{\varphi}_n$. The expression clarifies the dependence of it on components of unitary matrix $U$ and initial qubit state $\varphi$. As a consequence, we present a new type of limit theorems for the quantum random walk. In contrast with the de Moivre-Laplace limit theorem, our symmetric case implies that $X^{\varphi} _n /n$ converges weakly to a limit $Z^{\varphi}$ as $n \to \infty,$ where $Z^{\varphi}$ has a density $1 / \pi (1-x^2) \sqrt{1-2x^2}$ for $x \in (- 1/\sqrt{2}, 1/\sqrt{2})$. Moreover we discuss some known simulation results based on our limit theorems.

%\footnote[0]{
%{\it Abbr. title:} Limit theorems for the quantum random walk.
%}

\footnote[0]{ This work was partially supported by the Grant-in-Aid for Scientific Research (B) (No.12440024) of Japan Society of the Promotion of Science. 

}

\end{small}

\setcounter{equation}{0}
\section{Introduction.}
\newcommand{\U}{\bar{U}}

The classical random walk on the line is the motion of a particle which inhabits the set of integers. The particle moves at each time step either one unit to the left with probability $p$ or one unit to the right with probability $q=1-p.$ The directions of different steps are independent of each other. This classical random walk is often called the simple random walk. For general random walks on a countable space, there is a beautiful theory (see Spitzer \cite{s}). In the presnet paper, we consider quantum variations of the classical random walk and refer to such processes as quantum random walks. 

\par
Very recently quantum random walks have been widely investigated by a number of groups in connection with the quantum computing, for exmples, \cite{a,cfg,k,kns,mb,mr,nv,t,y}. For more general setting including quantum cellular automata, see Meyer \cite{m}. This paper is an extended version of our previous short letter with no proof (Konno \cite{k}).

\par
In Ambainis {\it et al.} \cite{a}, they gave two general ideas for analyzing quantum random walks. One is the path integral approach, the other is the Schr\"odinger approach. In this paper, we take the path integral approach, that is, the probability amplitude of a state for the quantum random walk is given as a combinatorial sum over all possible paths leading to that state. 

\par
The quantum random walk considered here is determined by $2 \times 2$ unitary matrix $U$ stated in the next section. The new points of this paper is to introduce 4 matrices, $P, Q, R$ and $S$ given by the unitary matrix $U$, to obtain a combinatorial expression for the characteristic function by using them, and to clarify the dependence of the $m$th moment and symmetry of distribution for the quantum random walk on the unitary matrix $U$ and initial {\it qubit} (quantum bit) state $\varphi$. Furthermore we give a new type of limit theorems for the quantum random walk by using our results. Our limit theorem shows that the behavior of quantum random walk is remarkably different from that of the classical ramdom walk. As a corollary, it reveals whether some simulation results already known are accurate or not.

\par
The rest of the paper is organized as follows. In Section 2, we introduce a definition of the quantum random walk and explain our results. Section 3 gives the characteristic function. In Section 4, we present a condition for symmetry of the distribution. Section 5 is devoted to a proof of the limit theorem. In Section 6, we consider the Hadamard walk case.
\par

%%%%%%%%%%%%%%%%%%%%%%%%%%%%%%%%%%%%
\setcounter{equation}{0}
%%%%%%%%%%%%%%%%%%%%%%%%%%%%%%%%%%%%

\section{Definition and results.}
The time evolution of the one-dimensional quantum random walk studied here is given by the following unitary matrix:
\begin{eqnarray*}
U=
\left[
\begin{array}{cc}
a & b \\
c & d
\end{array}
\right],
\end{eqnarray*}
\par\noindent
where $a,b,c,d \in {\bf C}$ and ${\bf C}$ is the set of the complex numbers. So we have $|a|^2 + |c|^2 =|b|^2 + |d|^2 =1, \> a \overline{c} + b \overline{d}=0, \> c= - \triangle \overline{b}$, and $d= \triangle \overline{a}$, where $\overline{z}$ is the complex conjugate of $z \in {\bf C}$ and $\triangle = \det U = ad - bc.$ We note that the unitarity of $U$ gives $|\triangle|=1.$
\par
The quantum random walk is a quantum generalization of the classical random walk in one dimension with an additional degree of freedom called the chirality. The chirality takes values {\it left} and {\it right}, and means the direction of the motion of the particle. The evolution of the quantum random walk is given by the following way. At each time step, if the particle has the left chirality, it moves one unit to the left, and if it has the right chirality, it moves one unit to the right. More precisely, we present the left and right chirality states as $\ket{L}= {}^t[1,0]$ and $\ket{R}= {}^t[0,1]$, where $t$ indicates the transposed operator. So the unitary matrix $U$ acts on two chirality states $\ket{L}$ and $\ket{R}$ as 
\begin{eqnarray*}
U\ket{L} = a\ket{L} + c\ket{R}, \qquad U\ket{R} = b\ket{L} + d\ket{R}.  
\end{eqnarray*}

%
%Figure 1 depicts the move in a quantum random walk where the chirality undergoe%s a unitary transformation.

%%%%%%%%%%%%%%%%%%%%%%%%%%%%%%%%%%%%%%%%%%%%%%%%%%%%%%%%%%%%%%
%\begin{figure}[htbp]
%\begin{center}
%\includegraphics[width=5cm,clip]{fig001-1.eps}
%\qquad
%\includegraphics[width=5cm,clip]{fig001-2.eps}
%\caption{The dynamics of the quantum random walk}
%\end{center}
%\end{figure}
%%%%%%%%%%%%%%%%%%%%%%%%%%%%%%%%%%%%%%%%%%%%%%%%%%%%%%%%%%%%%

\par
The study on the dependence of some important quantities (e.g., characteristic function, the $m$th moment, limit density) on initial qubit state is one of the essential parts, so we define the set of initial qubit states as follows:
\[
\Phi = \left\{ \varphi = {}^t [\alpha, \beta] \in {\bf C}^2
: |\alpha|^2 + |\beta|^2 =1 \right\}.
\]
From now on, we will give a precise definition of the quantum random walk $X_n ^{\varphi}$ at time $n$ starting from initial qubit state $\varphi \in \Phi$. First we decompose $U=P+Q$, where
\begin{eqnarray*}
P=
\left[
\begin{array}{cc}
a & b \\
0 & 0 
\end{array}
\right], 
\quad
Q=
\left[
\begin{array}{cc}
0 & 0 \\
c & d 
\end{array}
\right].
\end{eqnarray*}
The important point is that $P$ (resp. $Q$) represents that the particle moves to the left (resp. right). We define the $(4N+2) \times (4N+2)$ matrix by
\begin{eqnarray*}
\overline{U}_N =
\left[
\begin{array}{ccccccc}
0 & P & 0 & \dots & \dots & 0 & Q \\
Q & 0 & P & 0 & \dots & \dots & 0 \\
0 & Q & 0 & P & 0 &\dots &0\\
\vdots & \ddots & \ddots & \ddots & \ddots & \ddots & \vdots \\
0 & \dots & 0 & Q & 0 & P & 0\\
0 & \dots & \dots & 0 & Q & 0 & P\\
P & 0 & \dots & \dots & 0 & Q & 0
\end{array}
\right], 
\qquad
\hbox{with}
\qquad
0 = 
\left[
\begin{array}{cc}
0 & 0 \\
0 & 0 
\end{array}
\right].
\end{eqnarray*}
Then we see that $\overline{U}_N$ becomes also unitary matrix, since $P$ and $Q$ satisfy
\begin{eqnarray*}
PP^*+QQ^* = P^* P + Q^*Q=
\left[
\begin{array}{cc}
1 & 0 \\
0 & 1  
\end{array}
\right]
, \quad PQ^*=QP^*=Q^*P=P^*Q=
\left[
\begin{array}{cc}
0 & 0 \\
0 & 0  
\end{array}
\right],
\end{eqnarray*}
where $*$ means the adjoint operator. Here an initial qubit state is given by 
\[
\Psi^{(0)} (\varphi) = 
\left[
\begin{array}{c}
0_N \\
\varphi \\
0_N 
\end{array}
\right]
\in {\bf C}^{4N+2}, 
\]
where $0_N = {}^t [0, \ldots, 0] \in {\bf C}^{2N}$ is the zero vector and $\varphi = {}^t [\alpha, \beta].$ The qubit state at time $n$, $\Psi ^{(n)} (\varphi)$, is determined by
\begin{eqnarray}
\Psi ^{(n)} (\varphi) = (\overline{U}_N)^n \Psi ^{(0)} (\varphi).
\label{eqn:eva}
\end{eqnarray}
We write
\begin{eqnarray*}
&&
\Psi^{(n)} (\varphi) = {}^t [\Psi_{-N} ^{(n)} (\varphi), \Psi_{-(N-1)} ^{(n)} (\varphi), \ldots, \Psi_{N} ^{(n)} (\varphi)],
\\
&&
\Psi_k ^{(n)} (\varphi) = 
\left[
\begin{array}{cc}
\Psi_{L,k} ^{(n)} (\varphi) \\
\Psi_{R,k} ^{(n)} (\varphi)   
\end{array}
\right] = 
\Psi_{L,k} ^{(n)} (\varphi) \ket{L} 
+ \Psi_{R,k} ^{(n)} (\varphi) \ket{R} 
\in {\bf C}^2.
\end{eqnarray*}
Then $\Psi_k ^{(n)} (\varphi)$ is the two component vector of amplitudes of the particle being at site $k$ and at time $n$ with the chirality being left (upper component) and right (lower component). We see that (\ref{eqn:eva}) implies
\begin{eqnarray}
\Psi_k ^{(n+1)} (\varphi) = 
(\overline{U}_N \Psi^{(n)} (\varphi))_k = Q \Psi_{k-1} ^{(n)} (\varphi) + P 
\Psi_{k+1} ^{(n)} (\varphi).
\label{eqn:kankei}
\end{eqnarray}
\par
We define the quantum random walk $X_n ^{\varphi}$ with state space $\{-N, \ldots, N \}$ by
\[
P(X_n ^{\varphi} = k) 
= \| \Psi_k ^{(n)} (\varphi) \|^2.
\]
By construction $P(X_0 ^{\varphi}=0)=1.$ We remark $P(X_n ^{\varphi} = k)$ is independent of $N$ as far as $n \le N.$  Hence we naturally regard $X_n ^{\varphi}$ as a ${\bf Z}$-valued random variable, where ${\bf Z}$ is the set of the integers, and denote by the same symbol $X_n ^{\varphi}.$ 
\par
In contrast with classical random walks, $X_n ^{\varphi}$ can not be written as $X_n ^{\varphi} = Y_1 + \cdots + Y_n,$ where $Y_1, Y_2, \ldots $ are independent and identically distributed random variables. It is also noted that the quantum random walk is not a stochastic process. It is a sequence of distributions arising from products of the unitary matrix $\overline{U}_N$. The unitarity of $\overline{U}_N$ ensures
\[
\sum_{k \in {\bf Z}}  P(X_n ^{\varphi} = k) 
= \| (\overline{U}_N) ^n \overline{\varphi} \|^2
= \| \overline{\varphi} \|^2
= |\alpha|^2 + |\beta|^2
= 1,
\]
for any $1 \le n \le N$ and initial state $\overline{\varphi}={}^t[0_N, \varphi , 0_N]$. That is, the amplitude always defines a probability distribution for the location. For initial state $\overline{\varphi}={}^t[0_N, \varphi , 0_N]$, we have 
\begin{eqnarray*}
&& \overline{U}_N \overline{\varphi} = {}^t[0_{N-1},P \varphi,0,Q \varphi,0_{N-1}], \\
&& \overline{U}_N ^2 \overline{\varphi} ={}^t[0_{N-2},P^2\varphi,0,(PQ+QP)\varphi,0,Q^2\varphi,0_{N-2}], \\
&& \overline{U}_N ^3 \overline{\varphi} = {}^t[0_{N-3},P^3\varphi,0,(P^2Q+PQP+QP^2)\varphi,0,(Q^2P+QPQ+PQ^2)\varphi,0,Q^3\varphi,0_{N-3}].
\end{eqnarray*}
This shows that expansion of $U^n = (P+Q)^n$ for the quantum random walk corresponds to that of $1^n = (p + q)^n$ for the classical random walk.

We now explain our results briefly. Using an explicit form of $P(X_n ^{\varphi}=k)$ (Lemma 3), we obtain the characteristic function of $X_n ^{\varphi}$ (Theorem 4) and the $m$th moment of it (Corollary 5). One of the interesting facts is that, when $m$ is even, the $m$th moment of $X_n ^{\varphi}$ is independent of the initial qubit state $\varphi \in \Phi$. On the other hand, when $m$ is odd, the $m$th moment depends on the initial qubit state. So the standard deviation of $X_n ^{\varphi}$ depends on the initial qubit state $\varphi \in \Phi$. Our main theorem is as follows:

\begin{thm}
\label{thm:thm1} Suppose $abcd \not= 0$. Then we have
\begin{eqnarray*}
\lim_{n \to \infty} {X_n ^{\varphi} \over n} = Z^{\varphi} \qquad 
\hbox{in law},
\end{eqnarray*}
where $Z^{\varphi}$ is a random variable whose distribution has a density 
$f_{\alpha, \beta} (x) dx$ such that 
\begin{eqnarray*}
f_{\alpha, \beta} (x)
= { \sqrt{1 - |a|^2} \> (1- \lambda_{\alpha, \beta} x) \over \pi (1 - x^2) \sqrt{|a|^2 - x^2}},
\qquad \hbox{with} \>\>\>
\lambda_{\alpha, \beta} =
|\alpha|^2 - |\beta|^2 + {a \alpha \overline{b \beta} + \overline{a \alpha} b \beta \over |a|^2 },
\end{eqnarray*}
for $x \in (- |a|, |a|)$, and $f_{\alpha, \beta} (x) = 0$ for $|x| \ge |a|.$ Here $\varphi={}^t[\alpha, \beta]$ as before.
\end{thm}

It can be confirmed that $f_{\alpha, \beta}(x)$ satisfies the property of a density function. Indeed, we see that $f_{\alpha, \beta}(x) \ge 0$ since $1 \pm \lambda_{\alpha, \beta} |a| \ge 0,$ and that 
\begin{eqnarray*}
\int_{-|a|} ^{|a|} f_{\alpha, \beta}(x) \> dx
&=&  { \sqrt{1 - |a|^2} \over \pi } \int_0 ^1 t^{-1/2}(1-t)^{-1/2}(1-|a|^2t)^{-1} \> dt 
\\
&=&  { \sqrt{1 - |a|^2} \over \pi } \Gamma (1/2)^2 {}_2F_1(1/2, 1; 1 ;|a|^2) 
\\
&=&
1.
\end{eqnarray*}
Here ${}_2F_1(a, b; c ;z)$ is the hypergeometric function (see Sect. 5). The last equality comes from $\Gamma (1/2)= \sqrt{\pi}$ and ${}_2F_1(1/2, 1; 1 ;|a|^2)=1/\sqrt{1-|a|^2}.$ Remark that 
\begin{eqnarray*}
E(Z^{\varphi}) 
= -  (1 - \sqrt{1 - |a|^2}) \> \lambda_{\alpha, \beta}, \quad 
E ((Z^{\varphi})^2) = 1 - \sqrt{1 - |a|^2}.
\end{eqnarray*}
Moreover, an easy computation shows that $| E((Z^{\varphi})^m)| \le 2 |a|^m$ for any $m \ge 1$. 

It should be noted that, if $|a|=1,$ then $b=c=0$ and $|d|=1$. So this case is trivial. In fact, Corollary 5 (ii) implies that $\lim_{n \to \infty} X_n ^{\varphi}/n = W^{\varphi}$, in law, where $W^{\varphi}$ is determined by $P(W^{\varphi} = -1) = |\alpha|^2$ and $P(W^{\varphi} = 1) = |\beta|^2$. Theorem 1 suggests the following result on symmetry of distribution for the quantum random walk (Theorem 6). Define
\begin{eqnarray*}
\Phi_s &=&  \{ \varphi \in 
\Phi : \> 
P(X_n ^{\varphi}=k) = P(X_n ^{\varphi}=-k) \>\> 
\hbox{for any} \> n \in {\bf Z}_+ \> \hbox{and} \> k \in {\bf Z}
\},
\\
\Phi_0 &=& \left\{ \varphi \in 
\Phi : \> 
E(X_n ^{\varphi})=0 \>\> \hbox{for any} \> n \in {\bf Z}_+
\right\},
\\
\Phi_{\bot} &=& \left\{ \varphi = {}^t[\alpha, \beta] \in 
\Phi :
|\alpha|= |\beta|=1/\sqrt{2}, \> a \alpha \overline{b \beta} + \overline{a \alpha} b \beta =0 
\right\}, 
\end{eqnarray*}
where ${\bf Z}_+$ is the set of the positive integers. For $\varphi \in \Phi_s$, the probability distribution of $X_n ^{\varphi}$ is symmetric for any $n \in {\bf Z}_+$. Using explicit forms of distribution of $X^{\varphi} _n$ (Lemma 3) and $E(X_n ^{\varphi})$ (Corollary 5 (i) for $m=1$ case), we have $\> \Phi_{s} = \Phi_0 = \Phi_{\bot}.$

%%%%%%%%%%%%%%%%%%%%%%%%%%%%%%%
\setcounter{equation}{0}
%%%%%%%%%%%%%%%%%%%%%%%%%%%%%%%

\section{Characteristic function.}
This section gives a combinatorial expression of the characteristic function of
 the quantum random walk $X^{\varphi} _n$. As a corollary, we obtain the $m$th moment of $X^{\varphi} _n$. For fixed $l$ and $m$, we consider
\[
\Xi(l,m)= \sum_{l_j, m_j \ge 0: l_1+ \cdots +l_n=l, m_1+ \cdots +m_n=m} P^{l_1}Q^{m_1}P^{l_2}Q^{m_2} \cdots P^{l_n}Q^{m_n}.
\]
It should be noted that for $l+m=n$ and $-l+m=k$, we see that $\Psi_k ^{(n)} (\varphi) = \Xi (l,m) \varphi,$ since $\Psi_k ^{(n)} (\varphi) = {}^t [\Psi_{L,k} ^{(n)} (\varphi), \Psi_{R,k} ^{(n)} (\varphi)] (\in {\bf C}^2)$ is a two component vector of amplitudes of the particle being at site $k$ at time $n$ for initial qubit state $\varphi \in \Phi$ and $\Xi (l,m)$ is the sum of all possible paths in the trajectory consisting of $l$ steps left and $m$ steps right with $l=(n-k)/2$ and $m=(n+k)/2$ (see (\ref{eqn:kankei})). For example, in the case of $P(X_4 ^{\varphi} = -2)$, we have the following expression:
%(see Figure 2):
\begin{eqnarray*}
\Xi (3,1) = QP^3 + PQP^2 + P^2QP + P^3Q.
\label{eqn:example}
\end{eqnarray*}

%%%%%%%%%%%%%%%%%%%%%%%%%%%%%%%%%%%%%%%%%%%%%%%%%%%%%%%%%%%%%
%\begin{figure}[htbp]
%\begin{center}
%\includegraphics[width=10cm,clip]{fig006.eps}
%\caption{Four different paths corresponding to $\Xi (3,1)$}
%\end{center}
%\end{figure}
%%%%%%%%%%%%%%%%%%%%%%%%%%%%%%%%%%%%%%%%%%%%%%%%%%%%%%%%%%%%%

\par\noindent
Here we find a nice relation: $P^2 = aP$. By using this, we have $\Xi (3,1) = a^2 QP + a PQP+ a PQP+ a^2 PQ.$ Moreover, to compute general $\Xi (l,m)$, it is convenient to introduce
\[
R=
\left[
\begin{array}{cc}
c & d \\
0 & 0 
\end{array}
\right], 
\quad
S=
\left[
\begin{array}{cc}
0 & 0 \\
a & b 
\end{array}
\right].
\]
Then we obtain the following table of products of the matrices $P,Q,R$ and $S$:
\par
\
\par
\begin{center}
\begin{tabular}{c|cccc}
  & $P$ & $Q$ & $R$ & $S$  \\ \hline
$P$ & $aP$ & $bR$ & $aR$ & $bP$  \\
$Q$ & $cS$ & $dQ$& $cQ$ & $dS$ \\
$R$ & $cP$ & $dR$& $cR$ & $dP$ \\
$S$ & $aS$ & $bQ$ & $aQ$ & $bS$ 
\end{tabular}
\end{center}
\begin{center}
Table 1
\end{center}
\par
\
\par\noindent
where $PQ=bR$, for example. Since $P, Q, R$ and $S$ form an orthonormal basis of the vector space of complex $2 \times 2$ matrices with respect to the trace inner product $\langle A | B \rangle = $ tr$(A^{\ast}B)$, $\Xi (l,m)$ has the following form:
\[
\Xi (l,m) = p_n (l,m) P + q_n (l,m) Q + r_n (l,m) R + s_n (l,m) S.
\]
Next problem is to obtain explicit forms of $p_n (l,m), q_n (l,m), r_n (l,m),$ and $s_n (l,m)$. The above example of $n=l+m=4$ case, we have
\begin{eqnarray*}
&& \Xi (4,0) = a^3 P, \quad
\Xi (3,1) = 2abc P + a^2b R + a^2c S, \quad \\
&& \Xi (2,2) =  bcd P + abc Q + b(ad+bc) R + c(ad+bc) S, \\
&& \Xi (1,3) = 2bcd Q + bd^2 R + cd^2 S, \quad
\Xi (2,2) = d^3 Q.
\end{eqnarray*}
So, for example, $p_4 (3,1)=2abc, \> q_4 (3,1)=0, \> r_4 (3,1)=a^2b,$ and $s_4 (3,1)=a^2c.$ The following holds in general.

\begin{lem}
\label{lem:lem2} We write $l \wedge m = \min \{l,m \}.$ Suppose $abcd \not= 0$. Then  
\par\noindent
\hbox{(i)} for $l \wedge m \ge 1$, we have 
\[
\Xi (l,m) = a^l \overline{a}^m \triangle^m 
\sum_{\gamma =1} ^{l \wedge m} \left(-{|b|^2 \over |a|^2} \right)^{\gamma} {l-1 \choose \gamma- 1} 
{m-1 \choose \gamma- 1} \left[ {l- \gamma \over a \gamma } P + {m - \gamma \over \triangle \overline{a} \gamma} Q - {1 \over \triangle \overline{b}} R + {1 \over b} S \right],
\]
\par\noindent
\hbox{(ii)} for $l \ge 1$ and $m = 0$, we have $\Xi (l,0) = a^{l-1} P,$
\par\noindent
\hbox{(iii)} for $l = 0$ and $m \ge 1$, we have $\Xi (0,m) = \triangle^{m-1} \overline{a}^{m-1} Q.$ 
\end{lem}

\begin{proof} We first calculate explicit forms of $p_n (l,m)$. To begin, we assume $l \ge 2$ and $m \ge 1$. From Table 1, it is sufficient to consider only the following case:
\begin{eqnarray*}
C(w)_{\gamma,l,m} = P^{w_1} Q^{w_2} P^{w_3} \cdots Q^{w_{2 \gamma}} P^{w_{2 \gamma+1}}, 
\end{eqnarray*}
where $w=(w_i) \in {\bf Z}_+ ^{2 \gamma +1},$ with $l= \sum_{k=0} ^{\gamma} w_{2k+1}$ and $m=\sum_{k=1} ^{\gamma} w_{2k} .$ For example, we take $w_1=w_2=w_3=1$ and $\gamma =1$ as $PQP$. We remark that $2 \gamma +1$ is the number of clusters of $P$'s and $Q$'s. Next we consider the range of $\gamma$. The minimum is $\gamma =1$, that is, 3 clusters. This case is
 $P \cdots P Q \cdots Q P \cdots P.$ The maximum is $\gamma = (l-1) \wedge m$. This case includes the patterns for example: 
\[
PQPQPQ \cdots PQPQ PP \cdots PP \>\> (l-1 \ge m), \quad PQPQPQ \cdots PQPQQ \cdots QQ P \>\> (l-1 \le m).
\]
We introduce a set of sequences with $2 \gamma +1$ components: for fixed $\gamma \in [1, (l-1) \wedge m]$, 
\begin{eqnarray*}
W_{\gamma,l,m}
= 
\{ w = (w_i) \in {\bf Z}_+ ^{2 \gamma +1} : 
\sum_{k=0} ^{\gamma} w_{2 k +1} =l, \> \sum_{k=1} ^{\gamma} w_{2 k} =m \}. 
\end{eqnarray*}
By a standard combinatorial argument, we have
\begin{eqnarray}
|W_{\gamma,l,m}| = { l-1 \choose \gamma } { m-1 \choose \gamma -1}.
\label{eqn:caseb}
\end{eqnarray}
Let $w \in W_{\gamma,l,m}.$ Then by using Table 1, we have
\begin{eqnarray}
C(w)_{\gamma,l,m}
&=& a^{w_1-1} P d^{w_2 -1} Q a^{w_3-1} P \cdots d^{w_{2 \gamma} -1} Q a^{w_{2 \gamma +1} -1} P \nonumber \\
&=& a^{l-(\gamma +1)} d^{m - \gamma} (PQ)^{\gamma} P \nonumber \\
&=& a^{l-(\gamma +1)} d^{m - \gamma} b^{\gamma} c^{\gamma} P .
\label{eqn:casea}
\end{eqnarray}
Combining (\ref{eqn:caseb}) and (\ref{eqn:casea}), we obtain
\begin{eqnarray*}
p_n (l,m) P
&=& \sum_{\gamma =1} ^{(l-1) \wedge m}
\sum_{w \in W_{\gamma,l,m} }
C(w)_{\gamma,l,m} 
\nonumber
\\ 
&=&
\sum_{\gamma =1} ^{(l-1) \wedge m}
{ l-1 \choose \gamma } { m-1 \choose \gamma -1} 
a^{l-(\gamma +1)}  b^{\gamma} c^{\gamma} d^{m - \gamma} P.
\label{eqn:sikip}
\end{eqnarray*}
When $l \ge 1$ and $m=0$, it is easy to see that $p_n (l,0) P = P^l = a^{l-1} P.$ Furthermore, when $l=1, m \ge 1$ and $l=0, m \ge 0$, it is clear that $p_n (l,m) = 0.$
\par
As in the case of $p_n (l,m)$, we compute $q_n (l,m), r_n (l,m)$, and $s_n (l,m)$ by considering the patterns $Q^{w_1} P^{w_2} Q^{w_3} \cdots P^{w_{2 \gamma}} Q^{w_{2 \gamma+1}}, \> P^{w_1} Q^{w_2} P^{w_3} \cdots Q^{w_{2 \gamma}},$ and $Q^{w_1} P^{w_2} Q^{w_3} \cdots P^{w_{2 \gamma}},$ respectively. Then we obtain 
\begin{eqnarray*}
&&
q_n (l,m) =
\left\{
\begin{array}{cl}
\sum_{\gamma =1} ^{l \wedge (m-1)} { l-1 \choose \gamma -1} { m-1 \choose \gamma } a^{l-\gamma}  b^{\gamma} c^{\gamma} d^{m - (\gamma +1)} 
& \mbox{for $l \ge 1, m \ge 2,$} \\
d^{m-1}
& \mbox{for $l = 0, m \ge 1,$} \\
0
& \mbox{for $m=1, l \ge 1$ and $\> m=0, l \ge 0,$}
\end{array}
\right.
\\
&&
r_n (l,m) =
\left\{
\begin{array}{cl}
\sum_{\gamma =1} ^{l \wedge m} { l-1 \choose \gamma -1} { m-1 \choose \gamma -1}a^{l-\gamma}  b^{\gamma} c^{\gamma -1} d^{m - \gamma} 
& \mbox{for $l, m \ge 1,$} \\
0
& \mbox{for $l \wedge m =0,$}
\end{array}
\right.
\\
&&
s_n (l,m) =
\left\{
\begin{array}{cl}
\sum_{\gamma =1} ^{l \wedge m} { l-1 \choose \gamma -1} { m-1 \choose \gamma -1}a^{l-\gamma}  b^{\gamma -1} c^{\gamma} d^{m - \gamma}
& \mbox{for $l, m \ge 1,$} \\
0
& \mbox{for $l \wedge m =0.$}
\end{array}
\right.
\end{eqnarray*}
For $l \wedge m \ge 1$, the above explicit forms of $p(l,m), q_n (l,m), r_n (l,m)$, and $s_n (l,m)$ imply
\begin{eqnarray*}
\Xi (l,m) =
a^l d^m \sum_{\gamma =1} ^{l \wedge m} \left( {bc \over ad} \right)^{\gamma}
{ l-1 \choose \gamma -1} { m-1 \choose \gamma -1}
\left[ {l- \gamma \over a \gamma } P + {m - \gamma \over d \gamma} Q + {1 \over c} R + {1 \over b} S \right].
\end{eqnarray*}
From $c= - \triangle \overline{b}$ and $d= \triangle \overline{a}$, the proof of Lemma 2 (i) is complete. Furthermore, parts (ii) and (iii) are easily shown, so we will omit the proofs of them. 
\end{proof}

The distribution of $X^{\varphi} _n$ can be derived from Lemma \ref{lem:lem2} by direct computation. Let $[x]$ denote the maximal integer smaller than or equal to $x$. Let
\begin{eqnarray*}
\kappa_{\gamma,\delta,n,k} =  {k-1 \choose \gamma- 1} 
{k-1 \choose \delta- 1} 
{n-k-1 \choose \gamma- 1} 
{n-k-1 \choose \delta- 1}. 
\end{eqnarray*}

\begin{lem}
\label{lem:lem3} For $k=1,2, \ldots , [n/2],$ we have
\begin{eqnarray*}
&&
P(X^{\varphi} _n=n-2k) \\
&&
\quad
=  |a|^{2(n-1)} 
\sum_{\gamma =1} ^{k} \sum_{\delta =1} ^{k}
\left(-{|b|^2 \over |a|^2} \right)^{\gamma + \delta} 
\left( { \kappa_{\gamma,\delta,n,k} \over \gamma \delta} \right) \>
\Biggl[ 
\{ k^2 |a|^2 + (n-k)^2|b|^2 - (\gamma + \delta) (n-k)\} |\alpha|^2 
\\
&&
\qquad \qquad \qquad \qquad \qquad
+
\{ k^2 |b|^2 + (n-k)^2|a|^2  - (\gamma + \delta) k \} |\beta|^2 
\\
&&
\qquad \qquad \qquad \qquad \qquad 
+ {1 \over |b|^2}
\biggl[ 
\{ (n-k) \gamma - k \delta + n(2k-n) |b|^2 \} a \alpha \overline{b \beta} 
\\
&&
\qquad \qquad \qquad \qquad \qquad \qquad 
+ 
\{ -k  \gamma +(n-k) \delta + n(2k-n) |b|^2 \} \overline{a \alpha} b \beta + 
\gamma \delta 
\biggr] \Biggr],
\\
&&
P(X^{\varphi} _n=-(n-2k)) \\
&&
\quad
=  |a|^{2(n-1)} 
\sum_{\gamma =1} ^{k} \sum_{\delta =1} ^{k}
\left(-{|b|^2 \over |a|^2} \right)^{\gamma + \delta} 
\left( {\kappa_{\gamma,\delta,n,k} \over \gamma \delta} \right) \>
\Biggl[ 
\{ k^2 |b|^2 + (n-k)^2|a|^2 - (\gamma + \delta) k \} |\alpha|^2 
\\
&&
\qquad \qquad \qquad \qquad \qquad
+
\{ k^2 |a|^2 + (n-k)^2|b|^2 - (\gamma + \delta) (n-k) \} |\beta|^2 
\\
&&
\qquad \qquad \qquad \qquad \qquad
+ {1 \over |b|^2}
\biggl[ 
\{ k \gamma - (n-k) \delta - n(2k-n) |b|^2 \} a \alpha \overline{b \beta} 
\\
&&
\qquad \qquad \qquad \qquad \qquad \qquad 
+ 
\{ -(n-k) \gamma + k \delta - n(2k-n) |b|^2 \} \overline{a \alpha} b \beta + 
\gamma \delta 
\biggr] \Biggr],
\\
&&
P(X^{\varphi} _n=n) = |a|^{2(n-1)} \{ |b|^2 |\alpha|^2 +|a|^2 |\beta|^2- (a \alpha \overline{b \beta} + \overline{a \alpha} b \beta ) \}, \\
&&
P(X^{\varphi} _n=-n) = |a|^{2(n-1)} \{ |a|^2 |\alpha|^2 +|b|^2 |\beta|^2+ (a \alpha \overline{b \beta} + \overline{a \alpha} b \beta ) \}. 
\end{eqnarray*}
\end{lem}

By using Lemma \ref{lem:lem3}, we obtain a combinatorial expression for the characteristic function of $X^{\varphi} _n$ as follows. This result will be used in order to obtain a limit theorem of $X^{\varphi} _n$. Let $\mu_{\alpha,\beta} = \left(|a|^2 - |b|^2 \right)  
\left(|\alpha|^2 - |\beta|^2 \right) 
+ 2 (a \alpha \overline{b \beta} + \overline{a \alpha} b \beta )$ and $\nu_{\gamma,\delta,n,k} = (n-k)^2 + k^2 - n (\gamma + \delta) + 2 \gamma \delta /|b|^2.$

\begin{thm}
\label{thm:thm4} 
\par\noindent
\hbox{(i)} Suppose $abcd \not= 0$. Then we have
\begin{eqnarray*}
E(e^{i \xi X_n ^{\varphi}}) 
&=& 
|a|^{2(n-1)} 
\Biggl[
 \cos (n \xi) 
-  i \mu_{\alpha,\beta} \sin (n \xi)
\\
&& +  
\sum_{k=1}^{\left[{n-1 \over 2}\right]}
\sum_{\gamma =1} ^{k} \sum_{\delta =1} ^{k}
\left(-{|b|^2 \over |a|^2} \right)^{\gamma + \delta} 
\left( {\kappa_{\gamma,\delta,n,k} \over \gamma \delta} \right) \> 
\biggl[ \nu_{\gamma,\delta,n,k} \> \cos ((n-2k) \xi) 
\\
&&
\qquad \qquad 
- (n-2k) 
\biggl\{ \mu_{\alpha,\beta} \> n + {\gamma +\delta \over 2 |b|^2}  
(|\alpha|^2 - |\beta|^2 - \mu_{\alpha,\beta} ) \biggr\}
i \sin ((n-2k) \xi)
\biggr] 
\\
&& 
+
I \left( {n \over 2}-\left[{n \over 2} \right] \right)
\times
\sum_{\gamma =1} ^{{n \over 2}} \sum_{\delta =1} ^{{n \over 2}}
\left(-{|b|^2 \over |a|^2} \right)^{\gamma + \delta} 
\left( {\kappa_{\gamma,\delta,n,n/2} \over 2 \gamma \delta} \right) 
\nu_{\gamma,\delta,n,n/2}
\Biggr],
\end{eqnarray*}
where $I(x)=1 \> (resp. \> =0)$ if $x=0 \> (resp. \> x \not= 0).$ 
\par\noindent
\hbox{(ii)}
Let $b=0$. Then we have
\[
E(e^{i \xi X_n ^{\varphi}}) 
= \cos ( n \xi ) + i (|\beta|^2 - |\alpha|^2) \sin (n \xi).
\]
\mbox{(iii)}
Let $a=0$. Then we have
\[
E(e^{i \xi X_n ^{\varphi}}) =
\left\{
\begin{array}{cl}
\cos \xi + i  (|\alpha|^2 - |\beta|^2) \sin \xi
& \mbox{if $n$ is odd,} \\
1 
& \mbox{if $n$ is even.}
\end{array}
\right.
\]
\end{thm}

\par\noindent
We should remark that the above expression of the characteristic function in part (i) is not uniquely determined. From this theorem, we have the $m$th moment of $X^{\varphi} _n$ in the standard fashion. The following result can be used in order to study symmetry of distribution of $X^{\varphi} _n$.

\begin{cor}
\label{cor:cor5} 
\par\noindent
\hbox{(i)} Suppose $abcd \not= 0$. When $m$ is odd, we have
\begin{eqnarray*}
E((X_n ^{\varphi}) ^m) 
&=& 
- |a|^{2(n-1)}
\Biggl[ \mu_{\alpha,\beta} \> n^m + \sum_{k=1}^{\left[{n-1 \over 2}\right]}
\sum_{\gamma =1} ^{k} \sum_{\delta =1} ^{k}
\left(-{|b|^2 \over |a|^2} \right)^{\gamma + \delta} \> {(n-2k)^{m+1} \> \kappa_{\gamma,\delta,n,k} \over  \gamma \delta} 
\\
&&
\qquad \qquad \qquad \qquad \qquad \qquad  \times
\biggl\{ 
\mu_{\alpha,\beta} \> n + {\gamma +\delta \over 2 |b|^2}  
(|\alpha|^2 - |\beta|^2 - \mu_{\alpha,\beta} ) 
\biggr\} \Biggr].
\end{eqnarray*}
\par\noindent
When $m$ is even, we have
\begin{eqnarray*}
E((X_n^{\varphi}) ^m) 
&=& |a|^{2(n-1)} 
\Biggl\{
n^m +
\sum_{k=1}^{\left[{n-1 \over 2}\right]}
\sum_{\gamma =1} ^{k} \sum_{\delta =1} ^{k}
\left(-{|b|^2 \over |a|^2} \right)^{\gamma + \delta} 
\> {(n-2k)^{m} \kappa_{\gamma,\delta,n,k} \> \nu_{\gamma,\delta,n,k} \over  \gamma \delta} 
\Biggr\}.
\end{eqnarray*}
\par\noindent
\hbox{(ii)}
Let $b=0$. Then we have 
\[
E((X_n^{\varphi}) ^m) =
\left\{
\begin{array}{cl}
n^m (|\beta|^2 - |\alpha|^2)
& \mbox{if $m$ is odd,} \\
n^m  
& \mbox{if $m$ is even.}
\end{array}
\right.
\]
\par\noindent
\mbox{(iii)}
Let $a=0$. Then we have
\[
E((X_n ^{\varphi})^m) =
\left\{
\begin{array}{cl}
|\alpha|^2 - |\beta|^2
& \mbox{if $n$ and $m$ are odd,} \\
1
& \mbox{if $n$ is odd and $m$ is even,} \\
0
& \mbox{if $n$ is even.}
\end{array}
\right.
\]
\end{cor}

For any case, when $m$ is even, $E((X_n ^{\varphi})^m)$ is independent of initial qubit state $\varphi$. Therefore a parity law of the $m$th moment can be derived from the above result. 

%%%%%%%%%%%%%%%%%%%%%%%%%%%%%%%%%%%%
\setcounter{equation}{0}
%%%%%%%%%%%%%%%%%%%%%%%%%%%%%%%%%%%%

\section{Symmetry of distribution.}
In this section, we give a necessary and sufficient condition for the symmetry of the distribution of $X_n ^{\varphi}$.

\begin{thm}
\label{thm:thm6} Let $\Phi_{s}, \Phi_0,$ and $\Phi_{\bot}$ be as in Section 2. Suppose $abcd \not= 0$. Then we have $\Phi_{s} = \Phi_0 = \Phi_{\bot}.$
\end{thm}

This is a generalization of the result given by \cite{kns} for the Hadamard walk introduced in Section 6. Nayak and Vishwanath \cite{nv} discussed the symmetry of distribution and showed that ${}^t[1/\sqrt{2},\>$ $\pm i/\sqrt{2}] \in \Phi_s$ for the Hadamard walk.

\begin{proof} (i) $\Phi_s \subset \Phi_0$. This is obvious by definition.
\par\noindent
(ii) $\Phi_0 \subset \Phi_{\bot}$. By Corollary 5 (i) with $m=1$, we see that  $E(X^{\varphi} _1)=E(X^{\varphi} _2)=0$ if and only if $\mu_{\alpha,\beta}=0.$  Then this implies that for $n \ge 3$, Corollary 5 (i) with $m=1$ can be rewritten as  
\begin{eqnarray*}
E(X_n ^{\varphi}) = 
- { |a|^{2(n-1)} (|\alpha|^2 - |\beta|^2 )  \over 2 |b|^2}
\sum_{k=1}^{\left[{n-1 \over 2}\right]}
\sum_{\gamma =1} ^{k} \sum_{\delta =1} ^{k}
\left(-{|b|^2 \over |a|^2} \right)^{\gamma + \delta} 
\> {(n-2k)^{2} (\gamma + \delta) \> 
\kappa_{\gamma,\delta,n,k} \over \gamma \delta}. 
\end{eqnarray*}
Therefore $E(X_n ^{\varphi})=0 \> (n \ge 3)$ gives $|\alpha|=|\beta|$. Combining $|\alpha|=|\beta|$ with $\mu_{\alpha,\beta}=0$, we have the desired result.
\par\noindent
(iii) $\Phi_{\bot} \subset \Phi_s$. We assume that $|\alpha|=|\beta|=1/\sqrt{2}$ and $a \alpha \overline{b \beta} + \overline{a \alpha} b \beta= 0.$ By using these and Lemma \ref{lem:lem3}, we see that for $k=1,2, \ldots, [n/2],$
\begin{eqnarray*}
P(X_n ^{\varphi}=n-2k)
= P(X_n ^{\varphi}=-(n-2k)) 
= {|a|^{2(n-1)} \over 2} \>
\sum_{\gamma =1} ^{k} \sum_{\delta =1} ^{k}
\left(-{|b|^2 \over |a|^2} \right)^{\gamma + \delta} \> 
{ \kappa_{\gamma,\delta,n,k} \> \nu_{\gamma,\delta,n,k} \over 
\gamma \delta },  
\end{eqnarray*}
and $P(X_n ^{\varphi}=n)=P(X_n ^{\varphi}=-n)=|a|^{2(n-1)}|\alpha|^2$. So the desired conclusion is obtained. 
\end{proof}

%%%%%%%%%%%%%%%%%%%%%%%%%%%%%%%%%%%%
\setcounter{equation}{0}
%%%%%%%%%%%%%%%%%%%%%%%%%%%%%%%%%%%%

\section{Proof of Theorem 1.}
\par\noindent
Let $P^{\nu, \mu} _n (x)$ denote the Jacobi polynomial. Then it is well known that $P^{\nu, \mu} _n (x)$ is orthogonal on $[-1,1]$ with respect to $(1-x)^{\nu}(1+x)^{\mu}$ with $\nu, \mu > -1$, and that the following relation holds:
\begin{eqnarray}
P^{\nu, \mu} _n (x) = {\Gamma (n + \nu + 1) \over \Gamma (n+1) \Gamma (\nu +1)}
{}_2F_1(- n, n + \nu + \mu +1; \nu +1 ;(1-x)/2),
\label{eqn:norio}
\end{eqnarray}
where $\Gamma (z)$ is the gamma function and ${}_2F_1(a, b; c ;z)$ is the hypergeometric function:
\[
{}_2F_1(a, b; c ;z) = \sum_{n=0}^{\infty} 
{\Gamma (a+n) \over \Gamma (a)} {\Gamma (b+n) \over \Gamma (b)} {\Gamma (c) \over \Gamma (c+n)} \cdot {z^n \over n!}.
\]
Let $\rho_{n,k,i} = P^{i,n-2k} _{k-1}(2|a|^2-1)$ for $i=0,1.$ Then we see that
\begin{eqnarray*}
\sum_{\gamma =1} ^{k}
\left(-{|b|^2 \over |a|^2} \right)^{\gamma -1}
{1 \over \gamma} 
{k-1 \choose \gamma- 1}  
{n-k-1 \choose \gamma- 1} 
&=&
{}_2F_1(-(k-1), -\{(n-k)-1\}; 2 ;-|b|^2/|a|^2)
\\
&=&
|a|^{-2(k-1)} {}_2F_1(-(k-1), n-k+1; 2 ; 1- |a|^2)
\\
&=&
{1 \over k} |a|^{-2(k-1)} \rho_{n,k,1}.
\end{eqnarray*}
The first equality is given by the definition of the hypergeometric function (see p.35 in \cite{p}). The second equality comes from the relation: ${}_2F_1(a, b; c ;z) = (1-z)^{-a} {}_2F_1(a, c-b; c ;z/(z-1)).$ The last equality follows from (\ref{eqn:norio}). In a similar way, we have
\[
\sum_{\gamma =1} ^{k}
\left(-{|b|^2 \over |a|^2} \right)^{\gamma -1}
{k-1 \choose \gamma- 1}  
{n-k-1 \choose \gamma- 1} 
= |a|^{-2(k-1)} \rho_{n,k,0}.
\]
By using the above relations and Theorem 4, we obtain the following asymptotics of characteristic function $E(e^{i \xi X^{\varphi}_n/n})$: 
\begin{lem}
\label{lem:lem7}
If $n \to \infty$ with $k/n=x \in (-(1-|a|)/2, (1+|a|)/2)$, then
\begin{eqnarray*}
&& E(e^{i \xi X_n ^{\varphi}/n}) 
\sim 
\sum_{k=1}^{\left[{n-1 \over 2}\right]} |a|^{2n - 4k -2}|b|^4 \\
&& \times 
\left[ 
\left\{ {2x^2-2x+1 \over x^2} \> \rho_{n,k,1}^2 - {2 \over x} \> \rho_{n,k,0} \rho_{n,k,1} + {2 \over |b|^2} \> \rho_{n,k,0}^2 \right\} \cos ((1-2x) \xi) 
\right.
\\
&& 
\left.
\qquad \qquad - \left( {1-2x \over x} \right) 
\biggl\{ { \mu_{\alpha, \beta} \over x} \> \rho_{n,k,1}^2 
+ {|\alpha|^2 - |\beta|^2 - \mu_{\alpha,\beta} \over |b|^2} \> \rho_{n,k,0} \> \rho_{n,k,1} \biggl\} 
i \sin ((1-2x) \xi)
\right],
\end{eqnarray*}
where $f(n) \sim g(n) $ means $f(n)/g(n) \to 1 \> (n \to \infty)$. 
\par
\end{lem}

Next we use an asymptotic result on the Jacobi polynomial $P^{\alpha + an, \beta +bn} _n (x)$ derived by Chen and Ismail \cite{c}. By using (2.16) in their paper with $\alpha \to 0$ or $1, a \to 0, \beta = b \to (1-2x)/x, x \to 2|a|^2-1$ and $\triangle \to 4(1-|a|^2)\{ (2x-1)^2-|a|^2 \}/x^2$, we have the following lemma. It should be noted that there are some minor errors in  (2.16) in that paper, for example, $\sqrt{(-\triangle)} \to \sqrt{(-\triangle)}^{\> -1}$.  

\begin{lem}
\label{lem:lem8}
If $n \to \infty$ with $k/n=x \in (-(1-|a|)/2, (1+|a|)/2)$, then
\begin{eqnarray*}
&& \rho_{n,k,0} \sim  
{ 2 |a|^{2k-n} \over \sqrt{\pi n \sqrt{- \Lambda}} } \cos (An+B),  \\
&& \rho_{n,k,1} \sim  
{2 |a|^{2k-n} \over \sqrt{\pi n \sqrt{- \Lambda}}} \sqrt{{x \over (1-x)(1-|a|^2)}} \cos (An+B+ \theta), 
\end{eqnarray*}
where $\Lambda = (1-|a|^2) \{(2x-1)^2-|a|^2 \}$, $A$ and $B$ are some constants (which are independent of $n$), and $\theta \in [0, \pi/2]$ is determined by $\cos \theta = \sqrt{(1-|a|^2)/4x(1-x)}$. 
\end{lem}

{\it Proof of Theorem 1.} From the Riemann-Lebesgue lemma and Lemmas \ref{lem:lem7} and \ref{lem:lem8}, we see that
\begin{eqnarray*}
\lim_{n \to \infty} E(e^{i \xi {X_n ^{\varphi} \over n}}) 
&=& 
{ 1-|a|^2 \over \pi} 
\int^{{1 \over 2}} _{{1 - |a| \over 2}} 
{ \cos ((1-2x) \xi)
- i \lambda_{\alpha,\beta} (1-2x) \> 
\sin ((1-2x) \xi) \over x (1-x) \sqrt{(|a|^2-1) (4x^2-4x+1-|a|^2)}} \> dx  
\\
&=&
{ \sqrt{1-|a|^2} \over \pi} 
\int^{|a|} _{-|a|} 
{ \cos (x \xi)
- i \lambda_{\alpha,\beta} \> x  \> 
\sin (x \xi) \over (1-x^2) \sqrt{|a|^2-x^2} } \> dx
\\
&=&
\int^{|a|} _{-|a|} 
{ \sqrt{1 - |a|^2} \> (1 - \lambda_{\alpha,\beta} x) \over \pi (1 - x^2) \sqrt{|a|^2 - x^2}} e^{i \xi x}
\> 
dx.
\end{eqnarray*}
Hence $X^{\varphi} _n/n$ converges weakly to the limit $Z^{\varphi}$. \qed

%%%%%%%%%%%%%%%%%%%%%%%%%%%%%%%%%%%%
\setcounter{equation}{0}
%%%%%%%%%%%%%%%%%%%%%%%%%%%%%%%%%%%%

\section{Hadamard walk case.}
\par\noindent
In this section, we focus on the Hadamard walk, which has been extensively investigated in the study of quantum random walks. The unitary matrix $U$ of the Hadamard walk is defined by the following Hadamard gate (see Nielsen and Chuang \cite{nc}):
\begin{eqnarray*}
U = 
{1 \over \sqrt{2}}
\left[
\begin{array}{cc}
1 & 1 \\
1 & -1 
\end{array}
\right].
\end{eqnarray*}
The dynamics of this walk corresponds to that of the symmetric random walk in the classical case. However the symmetry of the walk depends heavily on initial qubit state, see \cite{kns}. 

For example, in the case of the Hadamard walk with initial qubit state $\varphi = {}^t[1/\sqrt{2},i/\sqrt{2}]$ (symmetric case), direct computation gives 
\begin{eqnarray*}
&& P(X_4 ^{\varphi} = -4) = P(X_4 ^{\varphi} = 4) =1/16, \\
&& P(X_4 ^{\varphi} = -2) = P(X_4 ^{\varphi} = 2) =6/16, \quad 
P(X_4 ^{\varphi} = 0) = 2/16.
\end{eqnarray*}
In contrast with the above result, as for the classical symmetric random walk $Y^o _n$ starting from the origin, we see that
\begin{eqnarray*}
&& P(Y^o _4 = -4) = P(Y^o _4 = 4) =1/16, \\
&& P(Y^o _4 = -2) = P(Y^o _4 = 2) =4/16, \quad 
P(Y^o _4 = 0) = 6/16.
\end{eqnarray*}
In fact, quantum random walks behave quite differently from classical random walks. For the classical walk, the probability distribution is a binomial distribution. On the other hand, the probability distribution in the quantum random walk has a complicated and oscillatory form.

%Figure 3 shows symmetric probability distributions at time $n=100$ for both cla%ssical (dashed) and quantum (solid) random walks. In Figure 4, asymmetric proba%bility distribution for the quantum random walk st%arting from $\varphi = {}^t[%0,1]$ is presented.

%%%%%%%%%%%%%%%%%%%%%%%%%%%%%%%%%%%%%%%%%%%%%%%%%%%%%%%%%%%%
%\begin{figure}[htbp]
%\begin{tabular}{cc}
%\begin{minipage}{0.5\hsize}
%%\begin{center} 
%\includegraphics[width=7cm,clip]{fig002.eps}
%\caption{Symmetric distribution}
%\end{minipage}
%\begin{minipage}{0.5\hsize}
%\includegraphics[width=7cm,clip]{fig003.eps}
%\caption{Asymmetric distribution}
%\end{minipage}
%\end{tabular}
%\end{figure}
%%%%%%%%%%%%%%%%%%%%%%%%%%%%%%%%%%%%%%%%%%%%%%%%%%%%%%%%%%

Now we compare our analytical result (Theorem 1) with the numerical ones given by Mackay {\it et al.} \cite{mb}, Travaglione and Milburn \cite{t} for the Hadamard walk. In this case, Theorem 1 implies that for any initial qubit state $\varphi = {}^t[\alpha, \beta]$,
\begin{eqnarray*}
\lim_{n \to \infty} P(a \le X^{\varphi} _{n}/n \le b) = \int^b _a {  1-  (|\alpha|^2 - |\beta|^2 + \alpha \overline{\beta} + \overline{\alpha} \beta) x \over \pi (1-x^2) \sqrt{1-2x^2}} \> 1_{(-1/\sqrt{2},1/\sqrt{2})} (x) \> dx,
\end{eqnarray*}
where $1_{(u,v)}(x)$ is the indicator function, that is, $1_{(u,v)}(x)=1,$ if $x \in (u,v), \> = 0,$ if $x \notin (u,v).$ For the classical symmetric random walk $Y^o _n$ starting from the origin, the de Moivre-Laplace theorem shows 
\begin{eqnarray*}
\lim_{n \to \infty}
P(a \le Y^o _{n}/ \sqrt{n} \le b) = \int^b _a {e^{-x^2/2} \over \sqrt{2 \pi}} dx.
\end{eqnarray*}
If we take $\varphi = {}^t[1/\sqrt{2},i/\sqrt{2}]$ (symmetric case), then we have the following quantum version of the de Moivre-Laplace theorem: 
\begin{eqnarray*}
\lim_{n \to \infty} P(a \le X^{\varphi} _{n}/n \le b) = \int^b _a {1 \over \pi (1-x^2) \sqrt{1-2x^2}} \> 1_{(-1/\sqrt{2},1/\sqrt{2})} (x) \> dx.
\end{eqnarray*}

%%%%%%%%%%%%%%%%%%%%%%%%%%%%%%%%%%%%%%%%%%%%%%%%
%\begin{figure}[htbp]
%\begin{center}
%\includegraphics[width=5cm,clip]{fig004.eps}
%\end{center}
%\caption{Symmetric density function}
%\end{figure}
%%%%%%%%%%%%%%%%%%%%%%%%%%%%%%%%%%%%%%%%%%%%%%%%
%The symmetric limit density function is shown in Figure 5. 
\par\noindent
So there is a remarkable difference between the quantum random walk $X^{\varphi} _n$ and the classical one $Y^o _n$ even in a symmetric case. Noting that $E(X^{\varphi} _n)=0 \> (n \ge 0)$ for any $\varphi \in \Phi_{s}$, we have
\begin{eqnarray*}
\lim_{n \to \infty} sd(X_n ^{\varphi})/n =  \sqrt{(2 - \sqrt{2})/2} = 0.54119 \ldots , 
\end{eqnarray*}
where $sd(X)$ is the standard deviation of $X$. This rigorous result reveals that numerical simulation result 3/5 = 0.6 given by \cite{t} is not so accurate.
\par
As in a similar way, if we take $\varphi = {}^t[0,e^{i \theta}]$ with $\theta \in [0, 2\pi)$ (asymmetric case), then we see  
\begin{eqnarray*}
\lim_{n \to \infty} P(a \le X^{\varphi} _{n}/n \le b)  =  \int^b _a {1 \over \pi (1-x) \sqrt{1-2x^2}} \> 1_{(-1/\sqrt{2},1/\sqrt{2})} (x) \> dx.
\end{eqnarray*}

%%%%%%%%%%%%%%%%%%%%%%%%%%%%%%%%%%%%%%%%%%%%%%%%%%
%\begin{figure}[htbp]
%\begin{center}
%\includegraphics[width=5cm,clip]{fig005.eps}
%\end{center}
%\caption{Asymmetric density function}
%\end{figure}
%%%%%%%%%%%%%%%%%%%%%%%%%%%%%%%%%%%%%%%%%%%%%%%%%%

%See Figure 6 for the asymmetric limit density function. 
\par\noindent
So we have
\begin{eqnarray*}
\lim_{n \to \infty} E(X_n ^{\varphi})/n = (2 - \sqrt{2})/2 = 0.29289 \ldots , \quad 
\lim_{n \to \infty} sd(X_n ^{\varphi})/n = \sqrt{(\sqrt{2} -1)/2} = 0.45508 \ldots .
\end{eqnarray*}
When $\varphi = {}^t[0,1]$ ($\theta =0$), Nayak and Vishwanath \cite{nv} and
Ambainis {\it et al.} \cite{a} gave a similar result, but both papers did not treat weak convergence. The former paper took the Schr\"odinger approach, and the latter paper took two approaches, that is, the Schr\"odinger approach and the path integral approach. However both their results come mainly from the Schr\"odinger approach by using a Fourier analysis. The details on the derivation based on the path integral approach in [1] are not so clear compared with this paper.

\par
In another asymmetric case $\varphi = {}^t[e^{i \theta},0]$ with $\theta \in [0, 2\pi)$, a similar argument implies 
\begin{eqnarray*}
\lim_{n \to \infty} P(a \le X^{\varphi} _{n}/n \le b)  =  \int^b _a {1 \over \pi (1+x) \sqrt{1-2x^2}} \> 1_{(-1/\sqrt{2},1/\sqrt{2})} (x) \> dx.
\end{eqnarray*}
The symmetry of distribution gives the following same result as in the previous case $\varphi = {}^t[0, e^{i \theta}]$. So the standard deviation of the limit distribution $Z^{\varphi}$ is given by $\sqrt{(\sqrt{2} -1)/2} = 0.45508 \ldots.$ Simulation result 0.4544 $\pm$ 0.0012 in \cite{mb} (their case is $\theta =0$) is consistent with our rigorous result.
\par
\
\par\noindent
\par\noindent
{\bf Acknowledgment.} The author would like to thank the referee for the careful reading and useful suggestions which improve the paper.

\begin{small}

\bibliographystyle{plain}

\par
\vskip 1.0cm

Norio KONNO

Department of Applied Mathematics

Faculty of Engineering

Yokohama National University

Hodogaya-ku, Yokohama 240-8501

Japan

E-mail: norio@mathlab.sci.ynu.ac.jp

\end{small}

\end{document}